\documentclass[conference]{IEEEtran}
\IEEEoverridecommandlockouts
\usepackage{amsmath,amssymb,amsfonts}
\usepackage{algorithmic}
\usepackage{graphicx}
\usepackage{textcomp}
\usepackage{xcolor}
\usepackage{ulem} 
\usepackage{bm}
 
\def\BibTeX{{\rm B\kern-.05em{\sc i\kern-.025em b}\kern-.08em
    T\kern-.1667em\lower.7ex\hbox{E}\kern-.125emX}}
\begin{document}

\definecolor{lightgray}{gray}{0.9}
\definecolor{lightblue}{rgb}{0.8,0.9,1}
\newcommand{\riaz}{\textcolor{black}}
\newcommand{\matsumi}{\textcolor{black}}

\title{{\riaz{Accelerating HDC-CNN Hybrid Models Using Custom Instructions on RISC-V GPUs}}
{\footnotesize \textsuperscript{}}
\thanks{This study do not receive any particular fund}
}

\author{
    \IEEEauthorblockN{Wakuto Matsumi\textsuperscript{1st} and Riaz-Ul-Haque Mian\textsuperscript{2nd}}
    \IEEEauthorblockA{
        \textit{Shimane University, Interdisciplinary Faculty of Science and Engineering,} \\
        \textit{Information System Design and Data Science,} \\
        Matsue, Japan \\
        0009-0003-4617-2860, 0000-0001-6550-5753
    }
}

\maketitle

\begin{abstract}
\label{lbl:Abstract}
Accelerating HDC-CNN Hybrid Models on custom GPUs. Machine learning based on neural networks has been advancing rapidly. However, the energy consumption required for training and inference remains substantial, posing significant challenges in environments where energy efficiency is a critical concern. \textbf{Hyperdimensional Computing (HDC)} is an extremely lightweight machine learning approach designed to mimic the mechanisms of the human brain. While HDC exhibits a high degree of parallelism, its recognition accuracy for complex images is relatively low. To address this problem, accelerators that combine HDC and neural networks, such as Convolutional Neural Networks (CNN) have been proposed. However, their adoption is hindered by limitations in generalizability and programmability. Meanwhile, the emergence of \textbf{RISC-V} has accelerated the development of \textbf{open-source accelerators specialized for specific domains}. Traditionally, GPUs have been developed in a closed ecosystem, but there is a growing trend toward open-source GPU architectures, leading to the introduction of multiple \textbf{RISC-V-based GPUs}.  Like RISC-V CPU RISC-V GPU also have the potential to serve flexibility in custom HDC where conventional GPU are not design. Just as RISC-V CPUs have become the foundation for domain-specific accelerators, RISC-V GPUs have the potential to serve as a flexible and programmable platform, enabling efficient high-speed computation even in areas where conventional GPUs are less effective. In this study, we implement \textbf{custom instructions for GPU-based HDC operations}, establishing a computational framework that efficiently handles processing tasks using HDC. This approach enhances the applicability of HDC-CNN hybrid models, addressing the challenges of energy-efficient, high-performance computing. Our experimental analysis with four different custom instructions for HDC operations shows that it can achieve a speedup of up to 56.2 times when using a microbenchmark for bound operations.
\end{abstract}


\section{Introduction}
\label{lbl:Intro}
\begin{figure*}[t] 
    \centering
    \includegraphics[width=\textwidth]{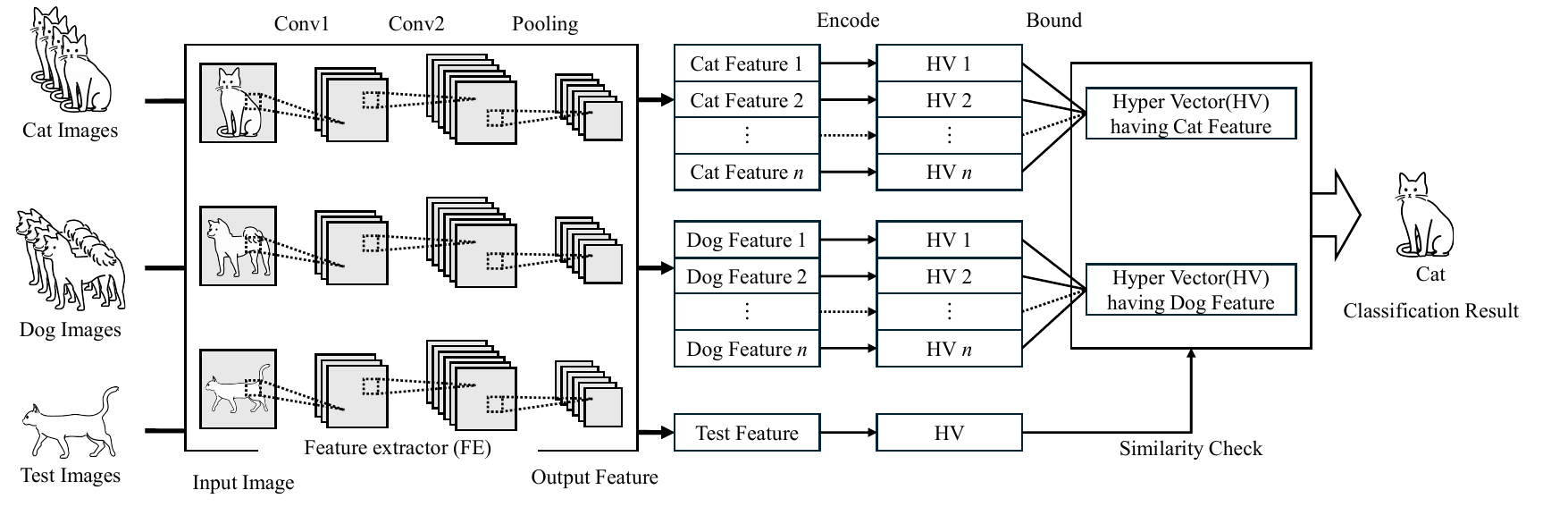}
    \caption{Overview of HDC-CNN Hybrid Models}
    \label{fig:hdc_cnn_hybrid}
\end{figure*}

{\riaz{Machine learning models based on neural networks have advanced significantly, but their high computational and energy costs pose challenges for efficient deployment. Hyperdimensional Computing (HDC) offers a lightweight, brain-inspired alternative that excels in parallelism and low power consumption. However, HDC alone struggles with complex image recognition tasks, leading to the development of HDC-CNN hybrid models that leverage Convolutional Neural Networks (CNNs) for improved accuracy. Despite these advancements, traditional GPUs face memory bottlenecks when processing HDC, limiting overall efficiency. In this work, we propose a custom GPU architecture based on RISC-V with specialized HDC instructions to accelerate HDC operations. The proposed approach enhances computational efficiency and enables seamless integration with CNNs, achieving significant speedup in hybrid model processing.}}

{\riaz{Hyperdimensional Computing (HDC) is an emerging lightweight machine learning paradigm inspired by the human brain. It encodes data into high-dimensional vectors, enabling massively parallel operations while reducing computational complexity and power consumption \cite{kanerva2009hyperdimensional}. Due to its efficiency, HDC is well-suited for edge computing applications. However, its recognition accuracy for complex datasets remains a challenge, leading to the development of HDC-CNN hybrid models that integrate Convolutional Neural Networks (CNNs) to enhance performance \cite{dutta_hdnn_pim}. Despite these advancements, executing HDC efficiently on traditional CPUs and GPUs is difficult due to their inherent architectural limitations. This necessitates the design of custom hardware accelerators that can efficiently process both HDC and hybrid models, motivating further research into specialized architectures \cite{isaka2024ecoflex}.}}


{\matsumi{Figure \ref{fig:hdc_cnn_hybrid} illustrates the processing flow of the HDC-CNN hybrid model.
The processing is divided into feature extraction by CNN and feature classification by HDC.
Existing CNNs such as VGG, ResNet, and MobileNet are used as feature extractors up to the first pooling layer.
HDC is used as the feature classifier.
The features output from the feature extractor are encoded into Hyper Vectors (HVs).
A Bound operation is performed on the encoded HVs for each classification class to create an HV representing the average feature.
The test data is also encoded into HV in the same way, and the class with the highest similarity to the HV representing the average feature is the classification result.}}

{\riaz{This research aims to propose a GPU-based platform capable of efficiently processing HDC by implementing custom instructions to support HDC operations on GPUs. This research will enable the high-speed execution of hybrid models combining HDC with other machine learning techniques, making it easier to validate related studies.}}

{\riaz{The key contributions of this paper are as follows:}}
\begin{itemize}
    \item We propose a \textbf{custom GPU architecture} based on RISC-V to efficiently accelerate Hyperdimensional Computing (HDC) operations, enabling seamless integration with Convolutional Neural Networks (CNNs) in HDC-CNN hybrid models.
    
    \item We introduce \textbf{four custom GPU instructions} tailored for HDC operations, significantly reducing memory access overhead and improving computational efficiency.
    
    \item We implement and evaluate the proposed architecture on the \textbf{Vortex RISC-V-based GPU}, demonstrating a maximum speedup of \textbf{56.2x} for microbenchmark-bound operations.
    
    \item We analyze the \textbf{performance bottlenecks} in HDC-CNN hybrid models and identify encoding as a major limiting factor, highlighting the need for further optimization in matrix operations.
    
    \item We provide a \textbf{comprehensive evaluation} comparing execution cycles before and after implementing custom GPU instructions, offering insights into their impact on real-world machine learning workloads.
\end{itemize}

{\riaz{The rest of the paper is organized as follows. Section~\ref{lbl:Background} provides the necessary background and discusses related work on Hyperdimensional Computing (HDC) and its integration with Convolutional Neural Networks (CNNs). Section~\ref{lbl:Preliminaries} introduces the fundamental concepts and computational techniques used in HDC, highlighting the key challenges in its hardware acceleration. Section~\ref{lbl:Proposed} presents the proposed custom GPU architecture, detailing the implementation of specialized HDC-specific instructions and optimization strategies. Section~\ref{lbl:Experiment} describes the experimental setup, evaluation methodology, and benchmark results, demonstrating the efficiency of our approach. Finally, Section~\ref{lbl:Conclusion} concludes the paper by summarizing the key findings and discussing potential future research directions.}}

\section{Background}
\label{lbl:Background}
{\riaz{To address the limitations of conventional neural network-based models in terms of computational efficiency and energy consumption, alternative learning paradigms and specialized architectures have been explored. The following sections discuss Hyperdimensional Computing (HDC) as a lightweight machine learning approach and the role of domain-specific architectures, particularly RISC-V-based GPUs, in accelerating such models.
}}
\subsection{Hyperdimensional Computing}

Neural network-based machine learning has been rapidly advancing. In particular, the progress in generative AI has been remarkable, and a development race fueled by massive computational resources is underway, based on scaling laws \cite{kaplan_scaling_2020}. The forward and backward propagation used in neural network training and inference can be represented as matrix operations. Therefore, achieving fast matrix operations is essential for the rapid training and inference of neural networks. GPUs, which can efficiently execute matrix operations in parallel and are also capable of handling general-purpose tasks, are well-suited for neural network processing and are extensively used in large-scale AI development.

While high-accuracy models using neural networks have been developed, these models often come with a large number of parameters, making them computationally intensive and challenging to deploy on edge devices. GPUs, which are capable of general-purpose parallel processing, suffer from low power efficiency, making them unsuitable for running high-accuracy, parameter-heavy models in power-constrained environments. To achieve higher power efficiency, many accelerators have been proposed, but their architectures differ from traditional processors, leaving challenges in terms of programmability.

One proposed solution to these challenges is Hyperdimensional Computing (HDC). HDC is a lightweight machine learning method inspired by the workings of the brain \cite{kanerva2009hyperdimensional}. Data is encoded into hypervectors (HVs), which are vectors with thousands to tens of thousands of dimensions. Classification tasks are processed through operations on these HVs. HV operations are independent across each element of the vector and consist of simple operations, enabling massive parallelism. Additionally, HDC requires fewer parameters and less computational power compared to neural networks. As a result, HDC can perform learning and inference with low power consumption, making it a promising candidate for edge computing applications with limited computational resources and power constraints.

{\riaz{While HDC offers high parallelism and energy efficiency, it struggles to scale with data complexity due to its shallow architecture and limited feature representation. As a result, a drawback of HDC is its low recognition accuracy for complex image datasets~\cite{dutta_hdnn_pim}.}} To address this, a hybrid approach combining Convolutional Neural Networks (CNNs) and HDC has been proposed \cite{dutta_hdnn_pim}. In this paper, we refer to this model as the HDC-CNN hybrid model. Combining neural networks with HDC has been shown to improve accuracy on complex datasets, and it is expected that research into hybrid models combining HDC with other techniques will continue to advance \cite{dutta_hdnn_pim, ma2024hyperdimensional}.

Another issue is that HDC is not well-suited for processing on CPUs or GPUs. HV operations involve computations using thousands to tens of thousands of bits, but CPUs lack sufficient parallelism, and GPUs face memory access bottlenecks \cite{datta2019programmable, kang_openhd}. To address this, research into dedicated accelerators for HDC is progressing \cite{isaka2024ecoflex}. However, this approach is insufficient when considering hybrid models like the HDC-CNN hybrid model, which combines HDC with other machine learning techniques. Using this approach for HDC-CNN hybrid models would require not only an HDC accelerator but also a GPU or CNN accelerator, making it difficult to achieve from the perspective of power consumption and hardware costs. \cite{dutta_hdnn_pim} proposes an in-memory computing-based accelerator for HDC-CNN hybrid models, which operates with very low power consumption and high speed, but it cannot process general neural networks other than CNNs, making it difficult to adapt to new techniques as they emerge. To solve these issues, a versatile system capable of efficiently handling multiple techniques is needed.

\subsection{Domain-Specific Architecture}

RISC-V is an open-source, royalty-free Instruction Set Architecture (ISA) that is being actively developed. It features a modular ISA design that allows for the implementation of only the necessary instructions, as well as a well-defined mechanism for adding custom instructions. This makes it possible to customize processors at low cost according to specific use cases, and it is expected to be widely adopted in fields such as AI and automotive~\cite{SHD2024RISC-V,Waterman2016RISCV,riaz1,riaz2,riaz3}.

{\riaz{
In recent years, the performance gains attributed to Moore's Law and Dennard scaling have decelerated. However, it is well-established that high performance and power efficiency can still be achieved through the implementation of Domain-Specific Architectures (DSAs) \cite{SHDGroup2024}. The capacity of RISC-V to facilitate the implementation of custom instructions renders it a particularly suitable foundation for DSAs \cite{Asanovic2016}. Notably, RISC-V possesses an open architecture not only at the Instruction Set Architecture (ISA) level but also in its processor implementations, a characteristic that distinguishes it from many contemporary architectures \cite{vortex_3d}.}}

Rocket Chip provides an interface called RoCC (Rocket Custom Coprocessor) between the core and accelerators. By using RoCC, developers can focus solely on designing the accelerator without needing to design the processor itself, making it easier to develop custom processors~\cite{recket_chip}. 

While GPUs are widely used as general-purpose accelerators, their development has traditionally been closed. With the emergence of RISC-V, several RISC-V-based GPUs have been proposed, and RISC-V GPUs are now being developed as open-source projects~\cite{vortex_3d,li2024ventus}. Just as RISC-V CPUs have been used as the foundation for accelerators, RISC-V GPUs could serve as a base for accelerators that maintain programmability and versatility while enabling high-speed computation in areas where traditional GPUs fall short.

\label{lbl:Preliminaries}

\section{Hyperdimensional Computing}
{\riaz{To understand the role of HDC in hybrid machine learning models, we begin by examining its theoretical structure and core computational principles.
}}
\subsubsection{Theory}
\begin{figure}
    \centering
    \includegraphics[width=0.99\linewidth]{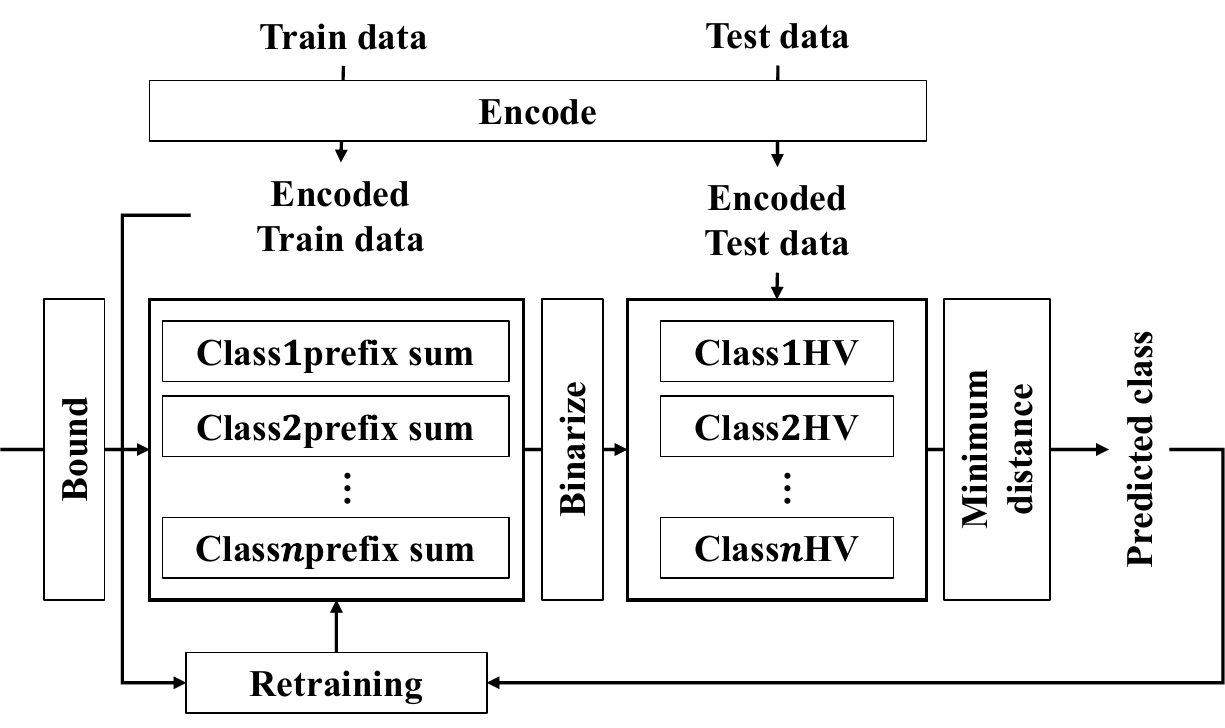}
    \caption{HDC Processing Workflow}
    \label{fig:hdc_flow}
\end{figure}

This section describes the classification task processing method using Hyperdimensional Computing (HDC).
The workflow of classification tasks in HDC is illustrated in Figure \ref{fig:hdc_flow}.

{\riaz{Classification tasks in HDC are divided into three stages:}}

\begin{enumerate}
    \item \textbf{Encoding:} Input data is converted into hypervectors (HVs).
    \item \textbf{Training:} Class HVs are generated based on relationships with multiple HVs.
    \item \textbf{Inference:} The similarity between encoded input HVs and class HVs is measured.
\end{enumerate}

Several variations of HVs are used in HDC, including real-valued vectors, complex vectors, bipolar vectors, and binary vectors.
HDC prioritizes vector dimensionality over individual element precision.
Due to this property, binary vectors are often preferred for their compatibility with computing architectures.

Depending on the characteristics of the target data, different encoding techniques are applied.
This section discusses two methods:
- Random Projection
- Locality-based Sparse Random Projection, which enhances computational efficiency \cite{imani_bric}.

Random Projection is a dimensionality reduction technique widely used in machine learning, not limited to HDC.
Let $\bm{F} = \{ f_1, f_2, ..., f_n \}, f_i \in \mathbb{N}$ be an $n$-dimensional input vector.
The encoded HV, $\bm{H} = \{ h_1, h_2, ..., h_D \}$, is derived using a random $D \times n$ projection matrix $\bm{P}$:

\[
h_i = sign(\bm{P}_i \cdot \bm{F})
\]

Thus, Random Projection transforms an $n$-dimensional input vector into a $D$-dimensional hypervector while preserving similarity relationships.

While Random Projection typically uses dense transformation matrices, it is possible to achieve similar accuracy using sparse matrices.
By carefully selecting non-zero elements in the transformation matrix, Locality-based Sparse Random Projection optimizes both computational complexity and memory efficiency.

If the sparsity factor is set to $s\%$, the non-zero elements in column $\bm{P}_i$ can be limited to $s \times n$, reducing memory storage and improving access patterns.
For these reasons, this paper adopts Locality-based Sparse Random Projection for encoding.

\subsubsection{Training Stage}
During the training phase, the system constructs class HVs representing different classification categories.
HDC learning is straightforward, where multiple HVs are aggregated to form class HVs.

Let $\bm{H}^1 = \{h^1_1, h^1_2, ..., h^1_D\}$ and $\bm{H}^2 = \{h^2_1, h^2_2, ..., h^2_D\}$ be two HVs belonging to the same class.
The class HV $\bm{H} = \{h_1, h_2, ..., h_D\}$ is obtained by majority voting across HV elements:

\begin{align*}
c_j &= \sum_i h^i_j \\
h_j &= sign(\frac{1}{2} + c_j)
\end{align*}

This process produces a representative HV for each class in HDC space.

\subsubsection{Inference Stage}
In the inference phase, the system compares the encoded HVs of test data against class HVs.
The similarity scores are computed for all class HVs, and the most similar class is assigned as the prediction result.

Although cosine similarity is commonly used, this paper adopts Hamming distance for computational efficiency.
The Hamming distance measures the number of differing elements between two HVs.
Since Hamming distance represents dissimilarity rather than similarity, the HV with the smallest distance is considered the most similar.

Retraining improves classification accuracy by updating class HVs based on prediction errors.
- If the predicted class matches the correct class, no updates are performed.
- If there is a misclassification, the incorrect HV is subtracted from the incorrect class sum and added to the correct class sum.

\begin{figure}
    \centering
    \includegraphics[width=0.9\linewidth]{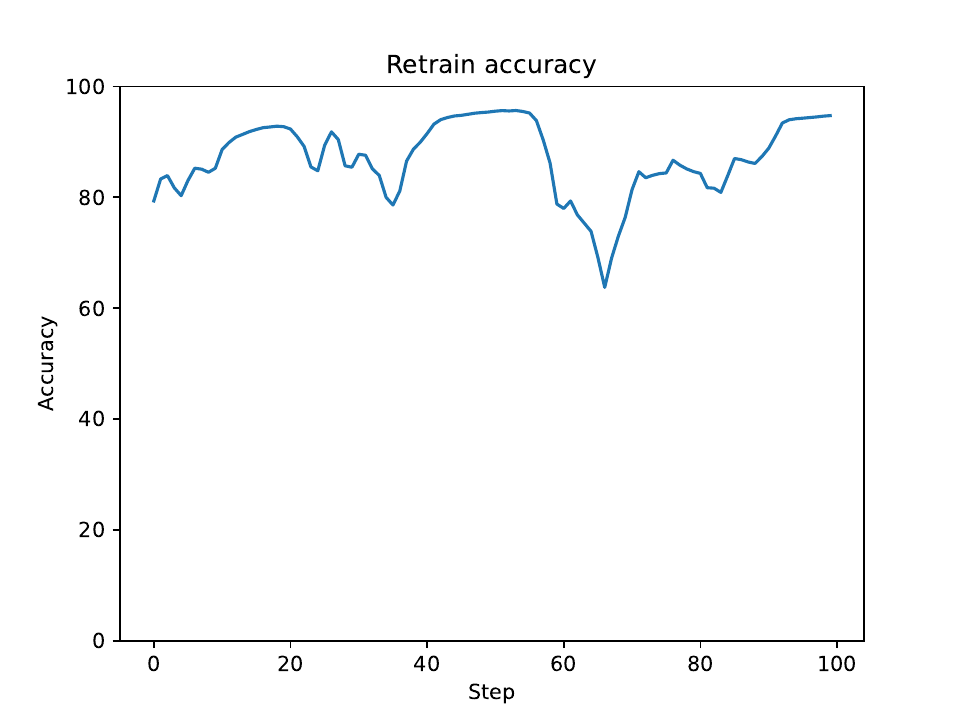}
    \caption{Accuracy Oscillation Due to Retraining}
    \label{fig:retrain_accuracy}
\end{figure}

This process adjusts the weight contributions of training data in forming class HVs.
However, as shown in Figure \ref{fig:retrain_accuracy}, retraining causes oscillations in accuracy.
Retraining is terminated once accuracy stabilizes.

\subsubsection{HDC-CNN}
HDC has applications in language recognition, speech recognition, and image recognition.
While HDC is lightweight and suitable for embedded environments, it struggles with complex image classification tasks such as CIFAR-10.

Dutta et al. proposed an HDC-CNN hybrid approach, where HDC is integrated with a CNN to achieve accuracy comparable to CNNs alone \cite{dutta_hdnn_pim}.
In this method, the output of the first CNN pooling layer is used as the input for HDC.
\cite{dutta_hdnn_pim} further developed an in-memory computing (PIM) accelerator to efficiently process both CNN and HDC computations.

However, dedicated HDC-CNN accelerators have limitations:
- They lack generalization beyond HDC-CNN models.
- Their architecture differs significantly from conventional processors, making programming more difficult.

Additionally, CPUs fail to fully exploit HDC and CNN parallelism, while GPUs face memory access bottlenecks, making it challenging to efficiently process both HDC and CNN on general-purpose processors.

\section{Proposed Method}
\label{lbl:Proposed}

\subsection{Proposed Accelerator and Custom GPU}

In the previous sections, we have identified the need for a \textbf{new accelerator} capable of executing various machine learning models efficiently while fully leveraging the \textbf{parallelism of Hyperdimensional Computing (HDC)}. 

To address this, our study proposes a \textbf{custom GPU architecture} that supports \textbf{HDC operations via custom instructions}, enabling efficient processing of both \textbf{HDC and CNN-based machine learning models}. 

The foundation of our \textbf{custom GPU implementation} is based on the \textbf{RISC-V ISA} and utilizes the \textbf{open-source Vortex GPGPU} \cite{vortex}, which has been actively developed for high-performance computing.

\subsection{Operating Theory}

The \textbf{custom instructions} are designed to reduce \textbf{memory access overhead} and minimize \textbf{computation cycles}, thereby improving processing speed. 

The optimization primarily targets:
\begin{itemize}
    \item The process of \textbf{deriving class HVs from pre-encoded training data}.
    \item The accumulation of \textbf{class-wise sums}, referred to as \textbf{Bound}.
    \item The derivation of class HVs from class sums, termed \textbf{Binarize}.
\end{itemize}

To \textbf{minimize memory usage}, each \textbf{hypervector (HV) element} is stored as a \textbf{single bit} in hardware. Additionally, we define \textbf{bipolar vectors}, where a bit value of $1$ represents $+1$ and a bit value of $0$ represents $-1$.

As discussed in the previous chapter, \textbf{Bound} refers to the \textbf{vertical accumulation sum} of HV elements. To align with \textbf{GPU register bit width}, we set the \textbf{counter width to 32-bit}. When performing \textbf{vertical accumulation} on a \textbf{32-bit register}, a \textbf{total of 32 counter elements} is required.

If this process were executed \textbf{without modifying the GPU}, relying solely on \textbf{memory and general-purpose registers}, \textbf{significant memory-to-register transfer overhead} would occur, leading to \textbf{slow processing}. In the \textbf{RISC-V-based GPU architecture}, each core has only \textbf{32 general-purpose 32-bit registers}, which is \textbf{insufficient} for handling all required counters. Consequently, \textbf{memory allocation for counter storage} becomes necessary, further increasing \textbf{memory access overhead} and slowing down computation.

To \textbf{resolve this issue}, we introduce \textbf{32 cumulative sum registers per thread}.  
These \textbf{cumulative registers}:
\begin{itemize}
    \item \textbf{Reduce dependency on memory bandwidth} by allowing \textbf{simultaneous retrieval of 32 counters}.
    \item \textbf{Minimize memory access operations}, limiting required reads to only \textbf{32-bit input data per cycle}.
    \item \textbf{Enable parallel execution} by implementing \textbf{32 arithmetic units per core}, ensuring \textbf{simultaneous computation of 32 elements}.
    \item \textbf{Achieve one-cycle execution per accumulation}, compared to \textbf{traditional architectures requiring 32 cycles}.
    \item \textbf{Optimize write-back efficiency}, allowing all \textbf{32 elements to be stored in memory simultaneously in one cycle}.
\end{itemize}

\begin{figure}[t]
  \centering
  \includegraphics[width=0.9\linewidth]{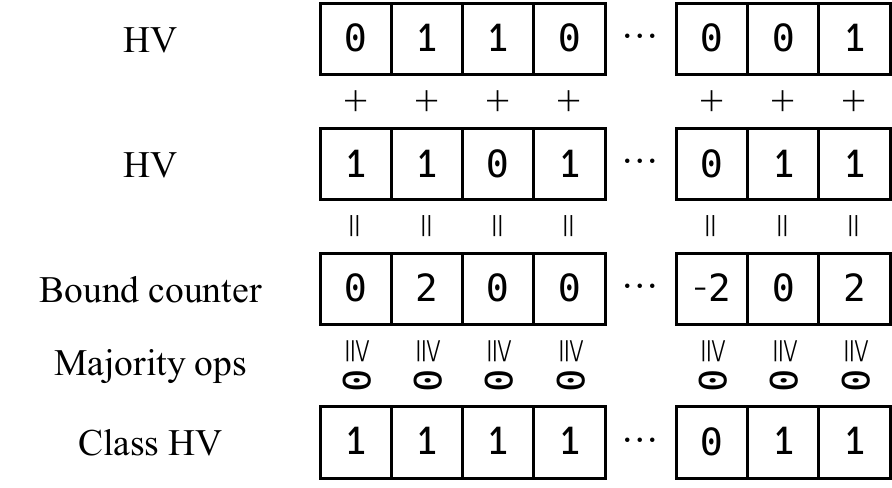}
  \caption{Parallel Execution Flow of Bound Computation Using Cumulative Sum Registers}
  \label{fig:bound}
\end{figure}

Additionally, during \textbf{Binarize processing}, \textbf{comparators are introduced to enable parallel execution}, ensuring \textbf{one-cycle Binarize computation}.  
Figure \ref{fig:bound} illustrates the \textbf{parallel execution flow of Bound operations using cumulative sum registers}, while Table \ref{tab:bound_cycle} compares the number of cycles required to derive class HVs with and without optimization.

\begin{table}[]
    \centering
    \caption{Cycle Count Comparison for Class HV Computation with $N$ 32-Element HVs}
    \begin{tabular}{|c||c|c|}
        \hline
                 & Conventional Method   & \emph{Proposed Method} \\ \hline \hline
         Input HV Loading & 1 $\times N$ & $N$ \\ \hline
         Counter Variable Read & 1 $\times$ 32 Elements $\times N$ & 0 \\ \hline
         Counter Variable Update & 1 $\times$ 32 Elements $\times N$ & $N$ \\ \hline
         Counter Variable Write-Back & 1 $\times$ 32 Elements $\times N$ & 0 \\ \hline
         Counter Variable Binarize & 2 $\times$ 32 Elements & 1 \\ \hline
         Total Cycle Count & $97N + 64$ & $2N + 1$ \\ \hline
    \end{tabular}
    \label{tab:bound_cycle}
\end{table}

\subsection{Implementation of Instructions}

To manipulate cumulative sum registers, we define \textbf{four R-type custom instructions}, as listed in Table \ref{tab:instr-format}.  
The \textbf{opcode} is assigned as \textbf{0x0b}, and \textbf{funct7} is set to \textbf{0x01}.

\begin{table}
    \centering
    \caption{Additional Custom Instructions}
    \label{tab:instr-format}
    \begin{tabular}{|c|c|c|}
        \hline
         Instruction Format & funct3 & Operation \\ \hline \hline
         vpopcnt.set rs1, rs2 & 1 & Assign value to Bound register \\ \hline
         vpopcnt.get rd, rs1 & 2 & Retrieve value from Bound register \\ \hline
         vpopcnt.add rs1 & 3 & Add bit array to Bound register \\ \hline
         vpopcnt.geq rd, rs1 & 5 & Compare Bound register value with threshold \\ \hline
    \end{tabular}
\end{table}

\section{Experiment and Result}
\label{lbl:Experiment}
\subsection{Experimental Setup}

We implemented all the instructions shown in Table \ref{tab:instr-format} in `simx`, the cycle-accurate simulator for the Vortex GPGPU. The benchmark programs were implemented as follows: the host-side program was written in C++, and the kernel-side program was implemented using OpenCL. Custom instructions were implemented using OpenCL's inline assembly.

The experimental environment is shown in Table \ref{tab:settings}. For the OpenCL implementation, we used PoCL, which is designed to target the Vortex GPGPU. The GPU core configuration was used without modification from its default settings.

\begin{table}
    \centering
    \caption{Experimental Environment}
    \label{tab:settings}
    \begin{tabular}{|c|c|}
        \hline
         Host CPU & Ryzen7 3700X \\ \hline
         Host Compiler & g++ 14.2.1 \\ \hline
         Device Compiler & Clang 18.1.7 \\ \hline
         OpenCL Implementation & PoCL \\ \hline
         OpenCL Version & 1.2 \\ \hline
         GPU Clusters & 1 \\ \hline
         GPU Cores & 1 \\ \hline
         GPU Warps & 2 \\ \hline
         GPU Threads & 4 \\ \hline
         GPU L1 Data Cache & 64KiB \\ \hline
         GPU L1 Instruction Cache & 16KiB \\ \hline
    \end{tabular}
\end{table}

The implemented custom instructions were evaluated using two benchmarks: a microbenchmark to measure the performance improvement of individual operations, and an image classification benchmark to measure performance in a practical scenario. For the microbenchmark, we applied the Bound operation to 1000 hypervectors (HVs) of 1024 dimensions. For the image classification benchmark, due to the simulation speed limitations, we extracted 5000 training images and 1000 test images from the MNIST dataset. The input images were encoded into 1024-dimensional HVs using Locality-based Random Projection, and the retraining process was fixed to 20 iterations. Measurements were taken using the Vortex GPGPU before and after implementing the custom instructions, and the performance was evaluated based on the ratio of execution cycles read from the GPU's internal counters. We chose to compare execution cycles because the simulator does not determine the GPU's clock frequency. The execution cycles counted only the cycles spent inside the OpenCL kernel, excluding the runtime execution during kernel launch and termination.

\subsection{Results}

The experimental results are shown in Table \ref{tab:spec_result}.

\begin{table}[]
    \centering
    \caption{Performance Improvement Ratio (in Cycles) of Custom GPU Over Baseline GPU}
    \begin{tabular}{|c|c|}
        \hline
        Microbenchmark (Bound) & 56.191x \\ \hline
        Image Classification Benchmark & 1.024x \\ \hline
    \end{tabular}
    \label{tab:spec_result}
\end{table}

The custom instructions were highly effective for the Bound operation microbenchmark, achieving a speedup of approximately 56x compared to the baseline. This improvement can be attributed to the reduction in the number of executed instructions due to parallelization, increased efficiency in instruction cache usage, and improved data cache utilization from counter registerization.

On the other hand, the impact of custom instructions on the image classification benchmark was limited. The primary reason for this limited effect is that the Bound operation accounted for only a small portion of the total processing time in this benchmark. The majority of the computation time in the image classification benchmark was dominated by the encoding process using Locality-based Random Projection. This process consists of matrix operations, which have lower parallelism compared to HDC operations like the Bound operation, making it the bottleneck. Due to the simulator's processing speed limitations, we did not perform CNN training on the same GPU in this experiment. If the combined processing with HDC involves large-scale operations like CNNs, the proportion of HDC operations in the total processing time of the hybrid model would be small, and the impact of HDC operation acceleration on reducing the overall processing time would likely be limited. However, optimizing the OpenCL kernel or introducing vector extension instructions \cite{riscvvector} from the RISC-V ISA used in this GPU could accelerate the matrix operations, which account for the largest portion of processing time, thereby contributing to overall performance improvement.

\section{Conclusion}
\label{lbl:Conclusion}

In this study, we proposed a computational framework that accelerates HDC operations by implementing custom instructions on a GPU, enabling efficient processing of hybrid models such as the HDC-CNN mixed model.

We implemented custom instructions supporting Bound operations in a simulator and conducted micro-benchmark evaluations, achieving a maximum speedup of approximately 56×.

However, in image classification benchmarks, the acceleration achieved was only about 2\%, indicating limited effectiveness of the custom instructions.  
The primary reason for the limited performance improvement in real-world applications is that the majority of processing time is spent on encoding, while the contribution of Bound operations to overall processing time is relatively small.  
Thus, the impact of Bound acceleration on total execution time remains limited.

To enhance encoding efficiency, optimization of matrix operations is necessary.  
This issue can potentially be addressed by optimizing OpenCL kernels and introducing extended instructions for matrix computations.

\section{Future Work}

Future challenges include:
- Evaluating circuit size and power consumption.  
- Measuring the overall performance of the HDC-CNN hybrid model.  
- Implementing additional custom instructions beyond Bound to support HDC computations.

Additionally, integrating optimized matrix computation techniques remains a key challenge for further accelerating overall processing.


\bibliographystyle{IEEEtran} 
\bibliography{bib}

@string{ISQED = "Proceedings of IEEE  International Symposium on Quality Electronic Design"}

@string{DATE = "Proceedings of IEEE Design Automation and Test in Europe"}

@string{APCCAS ="Proceedings of IEEE Asia Pacific Conference on Circuits and Systems"}

@string{CARRV ="Proceedings of Workshop on Computer Architecture Research with RISC-V"}

@string{SOCC ="Proceedings of IEEE International System-on-Chip Conference"}

@string{ICCD = "Proceedings of International Conference on Computer Design"}

@string{BERKELEY ="5-th Berkeley Symposium on Mathmatical Statistics and Probability"}

@article{recket_chip,
  title={The rocket chip generator},
  author={Asanovic, Krste and Avizienis, Rimas and Bachrach, Jonathan and Beamer, Scott and Biancolin, David and Celio, Christopher and Cook, Henry and Dabbelt, Daniel and Hauser, John and Izraelevitz, Adam and others},
  journal={EECS Department, University of California, Berkeley, Tech. Rep. UCB/EECS-2016-17},
  volume={4},
  pages={6--2},
  year={2016}
}

@phdthesis{Waterman2016RISCV,
  author       = {Andrew Waterman},
  title        = {Design of the RISC-V Instruction Set Architecture},
  school       = {University of California, Berkeley},
  year         = {2016},
  url          = {https://escholarship.org/uc/item/7zj0b3m7}
}

@inproceedings{vortex_3d,
  title={Vortex: Extending the RISC-V ISA for GPGPU and 3D-graphics},
  author={Tine, Blaise and Yalamarthy, Krishna Praveen and Elsabbagh, Fares and Hyesoon, Kim},
  booktitle={MICRO-54: 54th Annual IEEE/ACM International Symposium on Microarchitecture},
  pages={754--766},
  year={2021}
}

@inproceedings{vortex,
  title={Vortex risc-v gpgpu system: Extending the isa, synthesizing the microarchitecture, and modeling the software stack},
  author={Elsabbagh, Fares and Asgari, Bahar and Kim, Hyesoon and Yalamanchili, Sudhakar},
  booktitle={Third Workshop on Computer Architecture Research with RISC-V (CARRV 2019). Phoenix, AZ, USA,(June 22, 2019)},
  year={2019}
}

@article{datta2019programmable,
  title={A programmable hyper-dimensional processor architecture for human-centric IoT},
  author={Datta, Sohum and Antonio, Ryan AG and Ison, Aldrin RS and Rabaey, Jan M},
  journal={IEEE Journal on Emerging and Selected Topics in Circuits and Systems},
  volume={9},
  number={3},
  pages={439--452},
  year={2019},
  publisher={IEEE}
}

@inproceedings{ma2024hyperdimensional,
  title={Hyperdimensional computing vs. neural networks: Comparing architecture and learning process},
  author={Ma, Dongning and Hao, Cong and Jiao, Xun},
  booktitle={2024 25th International Symposium on Quality Electronic Design (ISQED)},
  pages={1--5},
  year={2024},
  organization={IEEE}
}

@article{kanerva2009hyperdimensional,
  title={Hyperdimensional computing: An introduction to computing in distributed representation with high-dimensional random vectors},
  author={Kanerva, Pentti},
  journal={Cognitive computation},
  volume={1},
  pages={139--159},
  year={2009},
  publisher={Springer}
}

@techreport{SHD2024RISC-V,
  author      = {The SHD Group},
  title       = {{RISC-V Market Report: Application Forecasts in a Heterogeneous World (Abridged Version)}},
  institution = {The SHD Group},
  year        = {2024},
  url         = {https://theshdgroup.com/wp-content/uploads/2024/01/RISC-V-Market-Analysis-2024-Abridged-Report-2.pdf}
}

@inproceedings{dutta_hdnn_pim,
	address = {Irvine CA USA},
	title = {{HDnn}-{PIM}: {Efficient} in {Memory} {Design} of {Hyperdimensional} {Computing} with {Feature} {Extraction}},
	isbn = {978-1-4503-9322-5},
	shorttitle = {{HDnn}-{PIM}},
	url = {https://dl.acm.org/doi/10.1145/3526241.3530331},
	doi = {10.1145/3526241.3530331},
	language = {en},
	urldate = {2023-12-18},
	booktitle = {Proceedings of the {Great} {Lakes} {Symposium} on {VLSI} 2022},
	publisher = {ACM},
	author = {Dutta, Arpan and Gupta, Saransh and Khaleghi, Behnam and Chandrasekaran, Rishikanth and Xu, Weihong and Rosing, Tajana},
	month = jun,
	year = {2022},
	pages = {281--286},
}

@inproceedings{isaka2024ecoflex,
  title={EcoFlex-HDP: High-Speed and Low-Power and Programmable Hyperdimensional-Computing Platform with CPU Co-Processing},
  author={Isaka, Yuya and Sakaguchi, Nau and Inoue, Michiko and Shintani, Michihiro},
  booktitle={2024 Design, Automation \& Test in Europe Conference \& Exhibition (DATE)},
  pages={1--6},
  year={2024},
  organization={IEEE}
}

@article{kang_openhd,
	title = {{OpenHD}: {A} {GPU}-{Powered} {Framework} for {Hyperdimensional} {Computing}},
	volume = {71},
	copyright = {https://ieeexplore.ieee.org/Xplorehelp/downloads/license-information/IEEE.html},
	issn = {0018-9340, 1557-9956, 2326-3814},
	shorttitle = {{OpenHD}},
	url = {https://ieeexplore.ieee.org/document/9785847/},
	doi = {10.1109/TC.2022.3179226},
	language = {en},
	number = {11},
	urldate = {2024-06-03},
	journal = {IEEE Transactions on Computers},
	author = {Kang, Jaeyoung and Khaleghi, Behnam and Rosing, Tajana and Kim, Yeseong},
	month = nov,
	year = {2022},
	pages = {2753--2765},
}

@inproceedings{imani_bric,
	address = {Las Vegas NV USA},
	title = {{BRIC}: {Locality}-based {Encoding} for {Energy}-{Efficient} {Brain}-{Inspired} {Hyperdimensional} {Computing}},
	isbn = {978-1-4503-6725-7},
	shorttitle = {{BRIC}},
	url = {https://dl.acm.org/doi/10.1145/3316781.3317785},
	doi = {10.1145/3316781.3317785},
	language = {en},
	urldate = {2024-12-26},
	booktitle = {Proceedings of the 56th {Annual} {Design} {Automation} {Conference} 2019},
	publisher = {ACM},
	author = {Imani, Mohsen and Morris, Justin and Messerly, John and Shu, Helen and Deng, Yaobang and Rosing, Tajana},
	month = jun,
	year = {2019},
	pages = {1--6},
}

@misc{kaplan_scaling_2020,
	title = {Scaling {Laws} for {Neural} {Language} {Models}},
	url = {http://arxiv.org/abs/2001.08361},
	doi = {10.48550/arXiv.2001.08361},
	language = {en},
	urldate = {2025-01-28},
	publisher = {arXiv},
	author = {Kaplan, Jared and McCandlish, Sam and Henighan, Tom and Brown, Tom B. and Chess, Benjamin and Child, Rewon and Gray, Scott and Radford, Alec and Wu, Jeffrey and Amodei, Dario},
	month = jan,
	year = {2020},
	note = {arXiv:2001.08361 [cs]},
}

@INPROCEEDINGS{li2024ventus,
  author={Li, Jingzhou and Yang, Kexiang and Jin, Chufeng and Liu, Xudong and Yang, Zexia and Yu, Fangfei and Shi, Yujie and Ma, Mingyuan and Kong, Li and Zhou, Jing and Wu, Hualin and He, Hu},
  booktitle={2024 IEEE 42nd International Conference on Computer Design (ICCD)}, 
  title={Ventus: A High-performance Open-source GPGPU Based on RISC-V and Its Vector Extension}, 
  year={2024},
  pages={276-279},
  doi={10.1109/ICCD63220.2024.00049}}

@misc{riscvvector,
  title        = "{RISC-V ”V” Vector Extension Version 1.0}",
  howpublished = "\url{https://github.com/riscvarchive/riscv-v-spec/releases/tag/v1.0}",
  author = {RISC-V International},
  note    = "[Accessed on 20-Jan-2025]"
}

@techreport{SHDGroup2024,
  author    = {T. S. Group},
  title     = {RISC-V Market Report: Application Forecasts in a Heterogeneous
               World (Abridged Version)},
  institution = {The SHD Group},
  year      = {2024},
  type      = {Tech. Rep.},
  url       = {https://theshdgroup.com/wp-content/uploads/2024/01/RISC-V-Market-Analysis-2024-Abridged-Report-2.pdf}
}

@techreport{Asanovic2016,
  author    = {Krste Asanovi\'c and Rimas Avizienis and John Bachrach and
               Scott Beamer and David Biancolin and Christopher Celio and
               Henry Cook and David Dabbelt and John Hauser and Adrian
               Izraelevitz and others},
  title     = {The rocket chip generator},
  institution = {EECS Department, University of California, Berkeley},
  year      = {2016},
  type      = {Tech. Rep.},
  number    = {UCB/EECS-2016-17},
  volume    = {4},
  pages     = {6--2}
}

@article{riaz1,
  title={Hardware--software co-design for decimal multiplication},
  author={Mian, Riaz-ul-haque and Shintani, Michihiro and Inoue, Michiko},
  journal={Computers},
  volume={10},
  number={2},
  pages={17},
  year={2021},
  publisher={MDPI}
}

@INPROCEEDINGS{riaz2,
  author={Riaz-ul-haque, Mian and Shintani, Michihiro and Inoue, Michiko},
  booktitle={2018 IEEE Asia Pacific Conference on Circuits and Systems (APCCAS)}, 
  title={Decimal Multiplication Using Combination of Software and Hardware}, 
  year={2018},
  volume={},
  number={},
  pages={239-242},
  doi={10.1109/APCCAS.2018.8605711}}

@INPROCEEDINGS{riaz3,
  author={Mian, Riaz-ul-haque and Shintani, Michihiro and Inoue, Michiko},
  booktitle={2019 32nd IEEE International System-on-Chip Conference (SOCC)}, 
  title={Cycle-Accurate Evaluation of Software-Hardware Co-Design of Decimal Computation in RISC-V Ecosystem}, 
  year={2019},
  volume={},
  number={},
  pages={412-417},
  doi={10.1109/SOCC46988.2019.1570559752}}

\vspace{12pt}

\end{document}